\documentclass[twoside,a4paper]{IEEEtran}

\usepackage{color}
\usepackage{graphicx}
\usepackage{epstopdf}
\usepackage{amsmath}
\usepackage{amssymb}
\usepackage{hyperref}
\usepackage[english]{babel}
\usepackage{cite}
\usepackage{rotfloat}
\usepackage{mathtools}
\usepackage[font=normalsize]{caption}
\usepackage{amsmath}
\usepackage{makecell}
\usepackage{algorithm,algorithmic}
\usepackage{multirow}
\usepackage{subfigure}
\usepackage{booktabs}
\usepackage{colortbl}
\usepackage{multirow}
\usepackage{hhline}
\usepackage{stfloats}
\usepackage{multicol}
\usepackage{upgreek}

\makeatletter

\def\ps@IEEEtitlepagestyle{%
	\def\@oddfoot{\mycopyrightnotice}%
	\def\@evenfoot{}%
}
\def\mycopyrightnotice{%
	{\footnotesize This work has been submitted to the IEEE for possible publication. Copyright may be transferred without notice, after which this version may no longer be accessible.\hfill}
	\gdef\mycopyrightnotice{}
}

\begin{document}

\title{\huge Hybrid Relay-Reflecting Intelligent Surface-Aided Wireless Communications: Opportunities, Challenges, \\ and Future Perspectives}

\author{Nhan~Thanh~Nguyen, Jiguang~He, Van-Dinh~Nguyen, Henk~Wymeersch, Derrick~Wing~Kwan~Ng, Robert~Schober, Symeon~Chatzinotas, and Markku~Juntti
		\thanks{\textit{N.~T.~Nguyen, J.~He, and M.~Juntti are with University of Oulu, Finland (E-mail: nhan.nguyen@oulu.fi, Jiguang.He@oulu.fi, markku.juntti@oulu.fi)}}
		\thanks{\textit{V.-D.~Nguyen and S. Chatzinotas are with University of Luxembourg (Email: dinh.nguyen@uni.lu, symeon.chatzinotas@uni.lu).}}
		\thanks{\textit{H.~Wymeersch is with Chalmers University of Technology, Sweden (E-mail: henkw@chalmers.se).}}
		\thanks{\textit{Derrick~Wing~Kwan~Ng is with University of New South Wales, Australia (E-mail: w.k.ng@unsw.edu.au).}}
		\thanks{\textit{Robert~Schober is with Friedrich-Alexander-University Erlangen-Nürnberg, Germany (E-mail: robert.schober@fau.de)}.}
		\vspace{-20pt}
}
\maketitle

\begin{abstract}
Reconfigurable intelligent surfaces (RISs) have emerged as a cost- and energy-efficient technology that can customize and program the physical propagation environment by reflecting radio waves in preferred directions. However, the purely passive reflection of RISs not only limits the end-to-end channel beamforming gains, but also hinders the acquisition of accurate channel state information for the phase control at RISs. In this paper, we provide an overview of a hybrid relay-reflecting intelligent surface (HR-RIS) architecture, in which only a few elements are active and connected to power amplifiers and radio frequency chains. The introduction of a small number of active elements enables a remarkable system performance improvement which can also compensate for losses due to hardware impairments such as the deployment of limited-resolution phase shifters. Particularly, the active processing facilitates efficient channel estimation and localization at HR-RISs. We present two practical architectures for HR-RISs, namely, fixed and dynamic HR-RISs, and discuss their applications to beamforming, channel estimation, and localization. The benefits, key challenges, and future research directions for HR-RIS-aided communications are also highlighted. Numerical results for an exemplary deployment scenario show that HR-RISs with only four active elements can attain up to $42.8$ percent and $41.8$ percent improvement in spectral efficiency and energy efficiency, respectively, compared with conventional RISs. 
\end{abstract} 

	
	
\IEEEpeerreviewmaketitle

\vspace{-10pt}
\section{Introduction}
Future wireless networks are expected to enable a fully connected, automated, and intelligent digital world by providing seamless and ubiquitous wireless connectivity with low cost and low complexity \cite{GiordaniCMag20}. With the recent breakthroughs in micro-electromechanical systems and meta-materials, it has become feasible to create smart, programmable, and reconfigurable radio environments via software-controlled reflection. The related technology is widely referred to as reconfigurable intelligent surfaces (RISs) \cite{RenzoJSAC20,KisseleffOJCOMS20}.


In practice, a RIS is typically composed of a large number of low-cost and low-energy consumption passive reflecting elements, each of which is able to reflect and control the incident wireless signals independently \cite{RenzoJSAC20,KisseleffOJCOMS20}. By judiciously adapting the phase shifts of the reflecting elements to the channel conditions, the reflected signals can be flexibly reconfigured to add constructively at the desired receivers, providing fine-grained and energy-focused reflective beamforming to improve wireless link performance and localization accuracy. Hence, it is not far-fetched to envisage that future wireless systems can be driven by the deployment of energy-neutral RISs to reduce the need for densely deployed, short-range, and power-hungry base stations (BSs).

Despite their huge potential and rich applications, recent studies showed that unless very large RISs are employed, they can be easily outperformed by conventional full-duplex (FD) amplify-and-forward (AF) relaying \cite{wu2019intelligent,bjornson2019intelligent}. On the other hand, when the size of a RIS is sufficiently large, additionally deploying a few more elements can only provide marginal performance improvement. This can be attributed to the finite-resolution RIS phase shifters, which restrict the promised performance, and the purely passive reflection, which limits the degrees of freedom for effective beamforming. These observations motivate a novel \emph{semi-passive beamforming} concept comprising both active processing and passive reflection to reap the key advantages of both RISs and relays.

In this paper, we present a \textit{hybrid relay-reflecting intelligent surface} (HR-RIS) architecture, which was originally proposed in~\cite{nguyen2021spectral,nguyen2021hybrid}. The main idea of HR-RISs is to connect a few reflecting elements to power amplifiers (PAs) and radio frequency (RF) chains, which are referred to as the \textit{active elements}. Although RISs with active sensors were introduced in \cite{taha2019deep} to facilitate channel estimation, they are unable to amplify the incident signals as HR-RISs. In practice, HR-RISs can be implemented by exploiting reflection amplifiers (RAs) \cite{landsberg2017low} based on low-power complementary metal-oxide-semiconductor (CMOS)-based technologies, which can flexibly alter the amplitude and phase shift of each element, facilitating higher beamforming gains and realizing robust HR-RIS-aided wireless systems. Furthermore, the active elements can be exploited for effective and accurate channel estimation. In fact, HR-RISs enable the design of sustainable and scalable wireless networks having affordable hardware cost and power consumption. This paper aims to provide an overview of HR-RISs, including their signal model, emerging architectures, and core functionalities. The excellent performance of HR-RISs is confirmed numerically for an exemplary deployment scenario. The main design and implementation challenges for deploying HR-RIS-aided wireless systems along with potential solutions are also highlighted to drive the development of future research.

\section{HR-RIS-assisted Wireless Systems}
\begin{figure}[t]
	\hspace{-0.45cm}
	\includegraphics[scale=0.34]{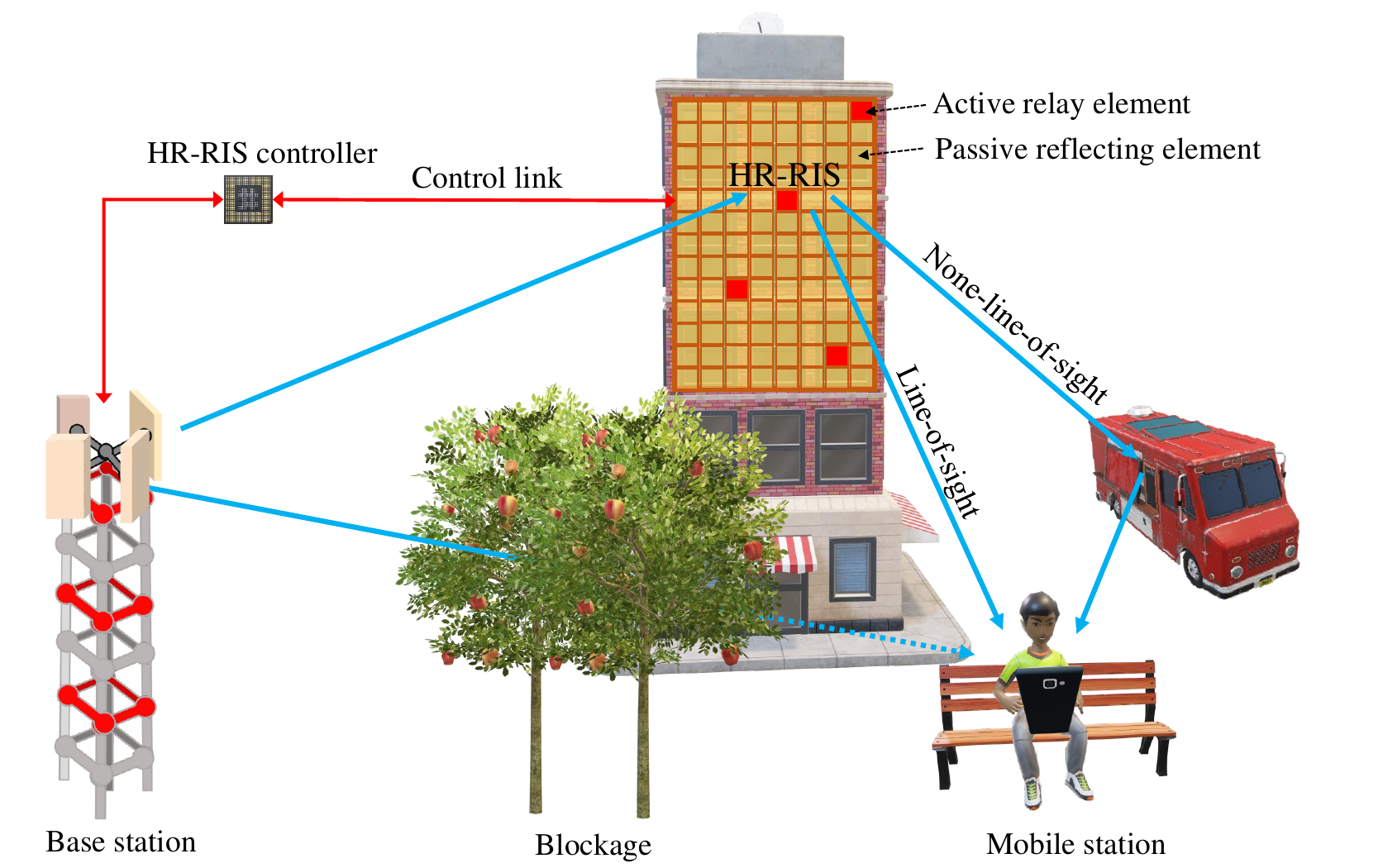}
	\caption{Illustration of an HR-RIS-aided wireless communication system.}
	\label{Fig_HR-RIS_system}
\end{figure}

Similar to conventional RISs, HR-RISs can be compactly and easily deployed on large surfaces to support different applications. For example, an HR-RIS may be mounted on a building to link a BS and a mobile station (MS), if the direct link between them is blocked by obstacles (e.g., buildings, trees), as depicted in Fig. \ref{Fig_HR-RIS_system}. In the following, we present the signal model of an HR-RIS-aided system, followed by two typical HR-RIS architectures.
\subsection{Signal Model}
\label{sec_system_model}
In an HR-RIS-assisted system, the composite channel from the BS to the MS is a concatenation of three components, including the link between the BS and the HR-RIS, the reflection/relaying at the HR-RIS, and the link between the HR-RIS and the MS. In general, HR-RISs play a similar role as conventional RISs by linearly mapping the impinging signals to output signals. However, the incident signals are not only reflected but also amplified by the active elements installed at the HR-RISs, which can be modeled as independent FD-AF relays \cite{landsberg2017low}. This linear mapping can be represented by a diagonal matrix, whose non-zero entries are complex numbers representing the reflecting and relaying coefficients associated with the passive and active elements, respectively. The amplitude and/or phase of a relaying/reflecting coefficient can be altered according to the dynamically varying propagation conditions for accurate beamforming. Specifically, the phases of both active and passive elements are tunable in the range $[0,2\pi)$. In general, the amplitude of a passive element is fixed to unity to maximize the received signal power and to enhance the channel capacity, as widely considered in the literature \cite{wu2019towards}. In contrast, the active elements of HR-RISs can also be optimized to amplify the incident signal for effective relaying. At the MS, the received signal can be modeled as the superposition of four components, including the relayed signals, noise, the self-interference (SI) caused by the active relay elements, and the reflected signals associated with the passive elements.

In practice, even if the optimal reflection coefficients of conventional RISs are applied, the system may not be able to guarantee a satisfactory achievable rate. This is due to the phenomenon of double path loss \cite{wu2019towards}, especially, when the signal is transmitted with low power and/or propagates through a channel with high path loss, such as in millimeter-wave (mmWave) and terahertz (THz) communications. In contrast, HR-RISs can overcome these limitations by reflecting signals in the preferred directions and adaptively amplifying them.

\subsection{HR-RIS Architectures}
\label{sec_architecture}
\begin{figure*}[t]
\centering
	\includegraphics[scale=0.55]{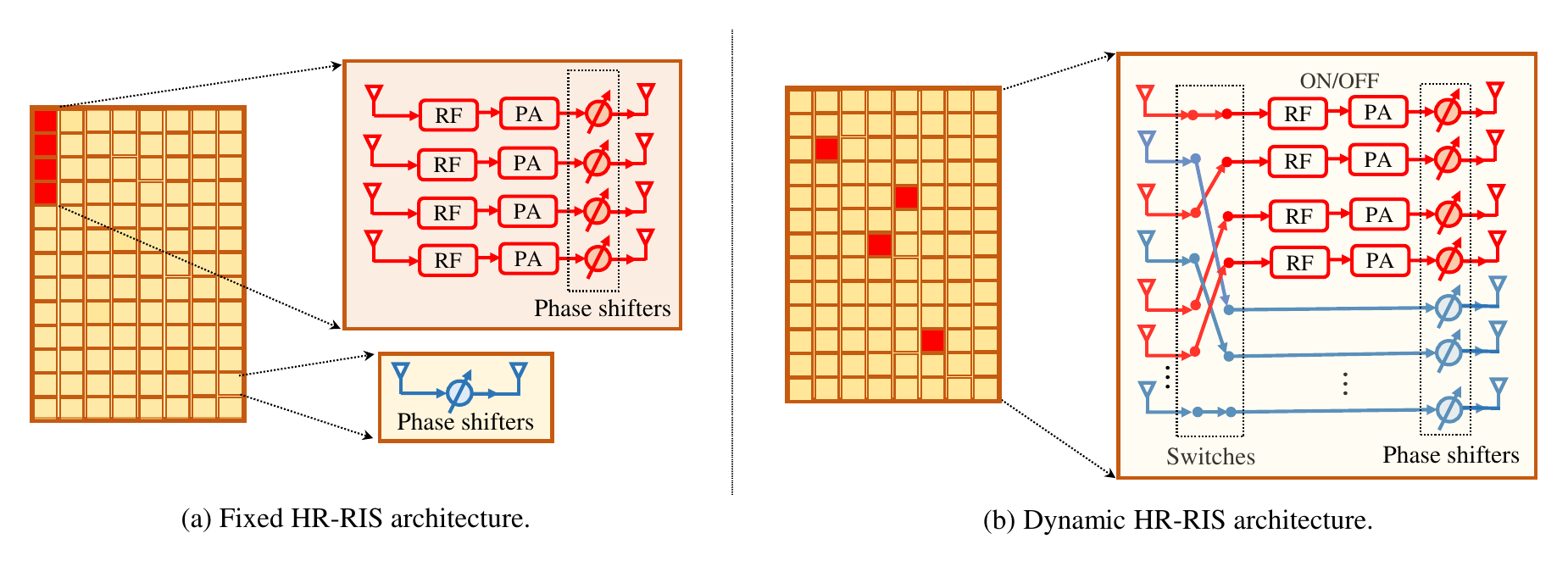}
	\caption{Illustration of (a) fixed and (b) dynamic HR-RIS architectures.}
	\label{fig_HR-RIS_architecture}
\end{figure*}
Compared with current RIS architectures, the most distinctive features of HR-RISs lie in the number and positions of the active elements \cite{nguyen2021hybrid}. In this regard, it is natural to either configure and fix them in advance or to dynamically optimize them based on the available channel state information (CSI). This corresponds to two different practical HR-RIS architectures (i.e., fixed and dynamic HR-RISs), which will be detailed next.

\subsubsection{Fixed HR-RIS}
This architecture is fixed in the sense that the number and positions of the active elements are predefined in the manufacturing process. As an example, a fixed HR-RIS with only four active elements located in the first column of a two-dimensional (2D) array is illustrated in Fig. \ref{fig_HR-RIS_architecture}(a). The received signal of an active element passes through the RF-PA chain and the associated phase shifter, while only the latter is needed for passive reflection. The fixed architecture is simple for implementation and the corresponding relaying coefficients can be obtained with low complexity. However, the diversity for active beamforming is not fully exploited due to the fixed configuration of the active elements. To overcome this limitation, a more sophisticated architecture, referred to as dynamic HR-RISs, is introduced.

\subsubsection{Dynamic HR-RIS}
In a dynamic HR-RIS, the number and positions of the active elements are dynamically determined based on the CSI and the power budget available for active processing. To this end, the elements in HR-RISs can be dynamically activated/deactivated to serve as active/passive elements, respectively. For example, an element should be activated only if the power amplification enabled by this element can improve the system performance (i.e., the amplitude of its optimized coefficient is larger than unity); otherwise, it is deactivated to serve as a passive reflecting element. As a result, dynamic HR-RISs can guarantee to provide at least the same beamforming gains compared to passive RISs of the same physical size. A dynamic HR-RIS is equipped with a few RF-PA chains, each of which can be turned on/off corresponding to the relaying/reflecting mode of the HR-RIS element \cite{nguyen2021hybrid}. More specifically, the elements of HR-RISs can dynamically connect to the RF-PA chains via a switching network as illustrated in Fig. \ref{fig_HR-RIS_architecture}(b).

Compared with fixed HR-RISs, dynamic HR-RISs can achieve higher active processing gains as the highly flexible architecture can adapt to the channel conditions for a better exploitation of the potential diversity gain. In addition, if some of the active elements are unable to amplify the incident signals due to hardware failure or limited power budget, they can be deactivated to serve as conventional passive reflecting elements, resulting in a higher passive beamforming gain and lower power consumption. However, these advantages of HR-RISs come at the cost of a more complicated architecture and more complex operation compared with their counterparts in fixed HR-RISs.

In practice, HR-RISs can be realized based on the recently introduced concept of low-power RAs \cite{landsberg2017low}. An RA of size $90$ $\upmu$m $\times$ $80$ $\upmu$m is capable of not only amplifying but also controlling the phase of an incident signal. The RA elements can operate in FD mode and SI can be suppressed solely in the analog domain to ensure low hardware cost \cite{landsberg2017low}. RAs can be printed on a surface, constructing an HR-RIS where all elements are active. An RA requires low power (e.g., $6$-$20$ mW \cite{landsberg2017low}), which, however, is still higher than that of a passive RIS element (e.g., $5$ mW \cite{bjornson2019intelligent}). Fortunately, HR-RISs can offer satisfactory relaying gains with only a single or a few active elements \cite{nguyen2021hybrid}. Therefore, a more practical and feasible implementation of fixed/dynamic HR-RISs is to upgrade conventional RISs by replacing a few/subset of elements with RAs, respectively, or by connecting them to low-power RF-PA chains \cite{nguyen2021hybrid}.

\section{Benefits of HR-RIS for Beamforming, Channel Estimation, and Localization}
\label{sec_application}

\begin{figure}[t]
	\centering
	\includegraphics[scale=0.32]{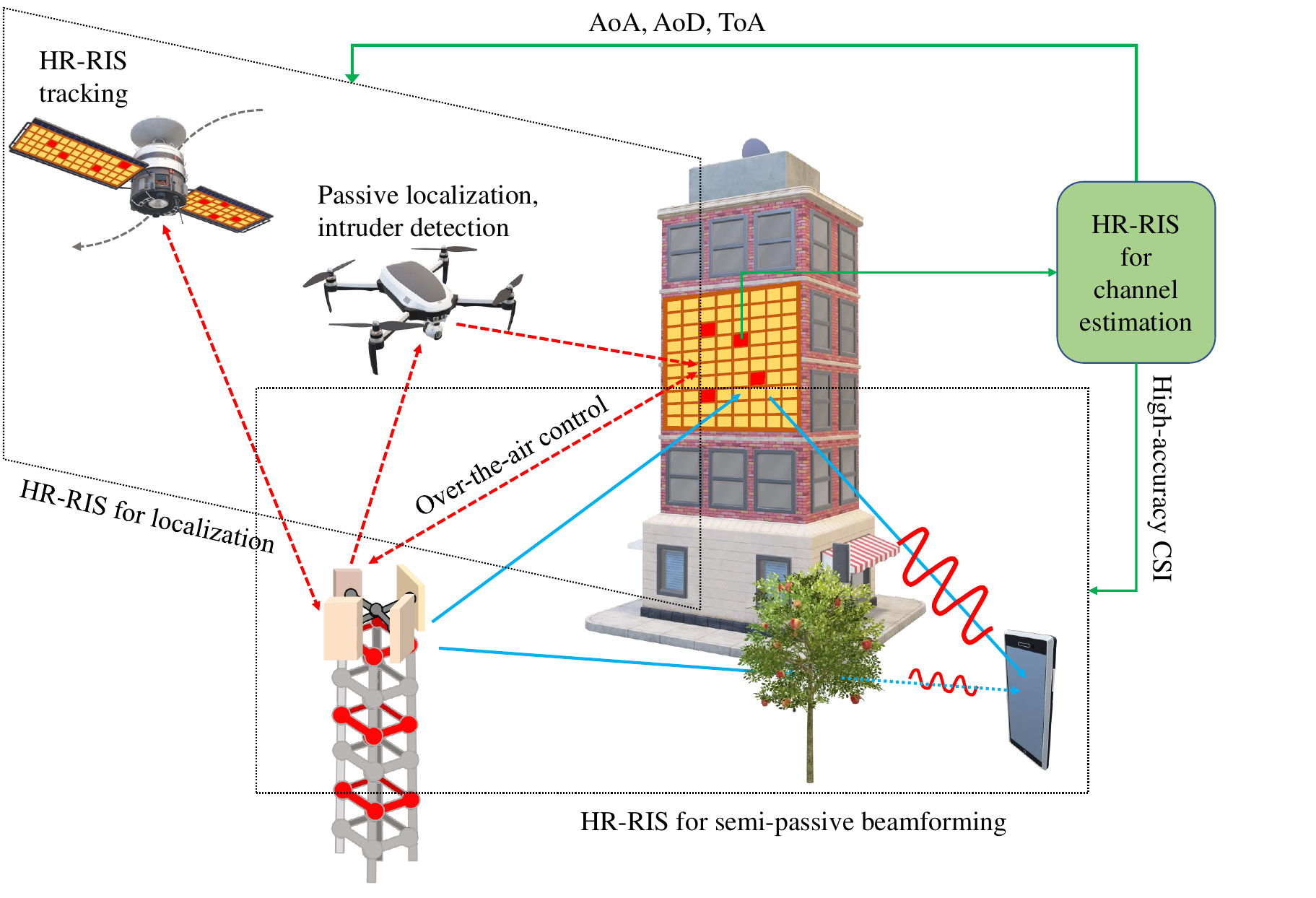}
	\caption{Application of HR-RIS to semi-passive beamforming, channel estimation, and localization.}
	\label{Fig_HRRIS_application}
\end{figure}

In Fig. \ref{Fig_HRRIS_application}, we illustrate the potential applications and benefits of HR-RISs. Specifically, they facilitate accurate channel estimation, and hence, are capable of improving the efficiency of beamforming and localization, which will be discussed in this section.

\subsection{Semi-Passive Beamforming}
\label{sec_beamforming}
The optimization of conventional RISs can be performed via designing a beamforming coefficient matrix with unity-magnitude diagonal elements \cite{wu2019towards, bjornson2019intelligent}. This is usually formulated as an optimization problem to determine the phase shifts of RISs that maximize the SE or the received signal strength. Although passive beamforming allows RISs to improve the system achievable rate, the resulting performance gains can be severely limited by various practical factors as discussed earlier. In contrast, HR-RISs enable semi-passive beamforming \cite{nguyen2021hybrid}, wherein both reflection and relaying are performed simultaneously.

Semi-passive beamforming enhances the performance of wireless systems, especially when the signal transmitted from the BS has low power and/or propagates through a channel with a severe propagation loss. In these cases, the power amplification provided by HR-RIS can substantially enhance the end-to-end channel conditions improving both the transmission rate and reliability. In general, the problem of semi-passive beamforming design inherits the mathematical challenges of conventional fully passive beamforming, (i.e., the non-convexity, high-dimension variables, etc.). It becomes even more challenging due to the additional constraints on the HR-RIS power budget as well as the number and distribution pattern of the active HR-RIS elements \cite{nguyen2021hybrid}. Despite these challenges, HR-RISs facilitate remarkable gains in both SE and energy efficiency (EE) compared with conventional RISs, as will be demonstrated later.

\subsection{Channel Estimation}
Conventional passive RISs are unable to process the received signals in the channel training phase. Therefore, the CSI can only be acquired at the BS or the MS via uplink or downlink pilot transmission, respectively \cite{He2020,he2020channel}. In this sense, the acquisition of individual large-dimension channels imposes great difficulties for the joint design of pilot sequences and RIS phase shift profiles.  As a remedy, the rank deficiency induced by the sparsity of the channel can be leveraged, and the channel parameters (e.g., the angles-of-departure (AoDs), angles-of-arrival (AoAs), and propagation path gains) can be extracted for effective channel estimation~\cite{he2020channel}. However, the high computational complexity and large training overhead required for channel estimation, especially for large RISs, remain challenging.

In HR-RIS-aided systems, channel estimation can be performed at HR-RISs by alternating between uplink and downlink training such that the channel estimation problem can be divided into two tractable subproblems, each mimicking a single point-to-point communication channel. Moreover, channel estimation for HR-RIS-aided systems can benefit from the channel sparsity and the rank deficiency by exploiting advanced compressive sensing approaches, similar to conventional RISs. In particular, the channel coefficients associated with the passive elements can be extracted by resorting to matrix completion techniques provided that some spatial correlation exists among the HR-RIS elements. Furthermore, thanks to their superior SE performance, HR-RISs generally can be deployed with much fewer elements compared with passive RISs, which significantly reduces the channel estimation complexity. Besides, the HR-RIS relaying gains generally lead to a higher received signal-to-noise ratio facilitating better channel estimation performance compared with the passive RISs.

\subsection{Localization}
Passive RISs can serve as references for positioning, leading to additional time-of-arrival (ToA), AoD, and AoA measurements \cite{wymeersch2020radio}. These measurements can be rendered resolvable from the uncontrolled multipath by optimizing RIS phase configurations to enhance localization accuracy and coverage. However, the purely passive processing of RISs makes it impossible to directly estimate the delays and angles of the propagation paths to/from RISs~\cite{bjornson2021reconfigurable}. 

The active processing in HR-RISs introduces additional benefits for localization. 
First of all, HR-RISs have a higher flexibility in terms of the design variables (phases and amplitudes), leading to a strictly better localization performance than passive RISs. In particular, new signal shapes in the spatial domain, inaccessible by passive RISs, can now be exploited by the active elements to significantly improve localization accuracy.
Second, an HR-RIS can act as a receiver and potentially combine the weighted observations from all the elements to determine the AoA of the impinging signal. Moreover, when the transmitter location is known, multipath measurements can be directly related to the objects in the environment, leading to an instantaneous three-dimensional mapping based on the AoA and ToA measurements. 
Third, by combining the received observations over time, an HR-RIS can determine its location and orientation, and feed back this information via its active RF chain through a dedicated control channel. This removes the need for careful placement and calibration as required for conventional RISs. Besides, assigning unique patterns (in terms of phase shift profiles) per HR-RIS provides a means to simultaneously determine the location and orientation of multiple HR-RISs, enabling applications with mobile HR-RISs (e.g., on a drone or a high-altitude platform system, as illustrated in Fig. \ref{Fig_HRRIS_application}). 
Fourth, the ability of HR-RISs to operate in receive mode provides a natural means for quickly conducting phase control without any feedback delay. This is particularly important for localization applications as they may be time-critical. 
The final important benefit is the potential to better exploit the wavefront curvature as HR-RISs can sample the incident wave, which can reduce the need for deploying expensive infrastructure.

\section{Comparison of HR-RISs, RISs, and Relays}
\label{sec_comparison}
Although HR-RISs, RISs, and relays share some similar functionalities, they have some fundamental differences, as will be discussed in the following.

\subsection{Hardware Complexity and Power Consumption}
HR-RISs require the connection of a few elements to RF-PA chains or the deployment of low-power RAs, causing additional power consumption and hardware cost for active processing. Fortunately, HR-RISs can be deployed with considerably reduced physical sizes compared with RISs. Therefore, the hardware complexity and the power consumption required for practical deployment of HR-RISs and RISs can be similar. Furthermore, the latency required for controling HR-RIS  is generally lower than that of the conventional RISs equipped with a large number of elements, thanks to the active elements equipped in the former.

To compare HR-RISs and relays, we recall that most of the elements in HR-RISs are passive and the remaining few active elements can be implemented with low-power RAs. These characteristics introduce the following properties. First, when a pure relay is equipped with as many elements as a typical HR-RIS (i.e., hundreds of elements), the former entails an excessively high and unaffordable hardware cost and power consumption. Second, a relay with a few elements may have a comparable hardware cost and power consumption as an HR-RIS but will achieve a lower performance as it cannot leverage passive reflection gains \cite{nguyen2021hybrid}.

\subsection{SE and EE Performance}
For sufficiently large-sized RISs, increasing the number of reflecting elements usually only yields a marginal additional passive beamforming gain. In contrast, introducing a few active elements can lead to a significant relaying gain \cite{bjornson2019intelligent, nguyen2021hybrid}. Hence, comparing a passive RIS with an HR-RIS of the same size, the latter has almost the same passive beamforming gain as the former but introduces substantial active relaying gains. This yields remarkable SE improvements, especially, when the signal is transmitted with low power and/or over a high path loss channel. Considering that only a small number of elements in HR-RISs are active, HR-RISs also provide improved EE compared with passive RISs.

Next, we compare the SE and EE performance of relays with that of HR-RISs and RISs. We note that a relay antenna can incur five and twenty times higher power consumption than that of an active RA and a passive reflecting element, respectively, deployed at RISs/HR-RISs~\cite{landsberg2017low ,bjornson2019intelligent}. Therefore, in practice, the number of elements at a relay is much smaller than those at a RIS and HR-RIS to ensure an affordable total power consumption. HR-RISs combine the advantages of such relays (i.e., small- and moderate-sized relays) and conventional RISs. Specifically, they not only attain a significant relaying gain, but also achieve additional passive reflection gains, comparable to those of conventional RISs. Therefore, HR-RISs are expected to outperform small- and moderate-sized relays in terms of both SE and EE, as will be numerically verified in the next section.

\section{Exemplary Deployment Scenario}
\label{sec_numerical_results}
\begin{figure}[t]
	\centering
	\subfigure[SE performance.]
	{
	    \includegraphics[scale=0.6]{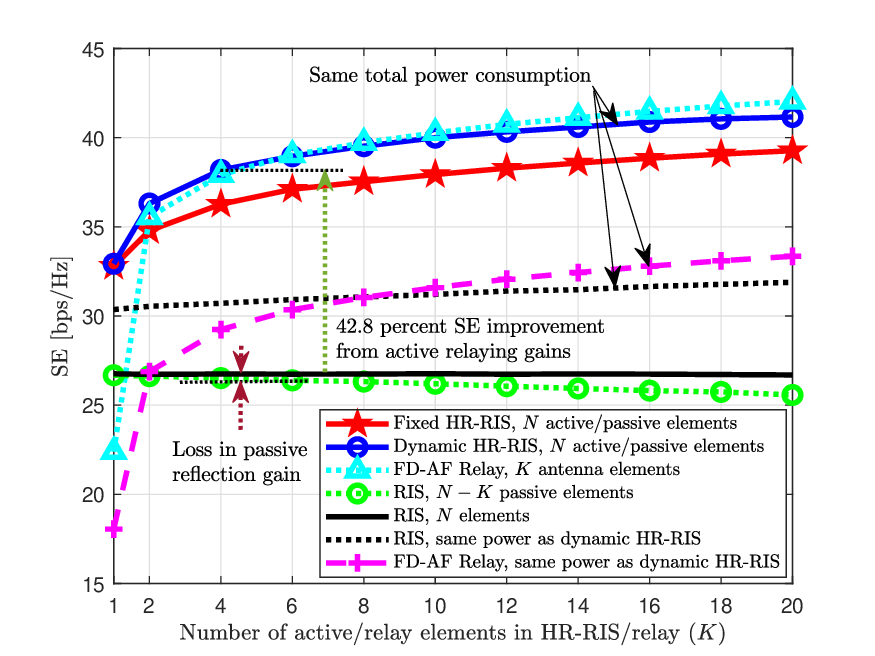}
    }
	\subfigure[EE performance.]
	{
	    \includegraphics[scale=0.6]{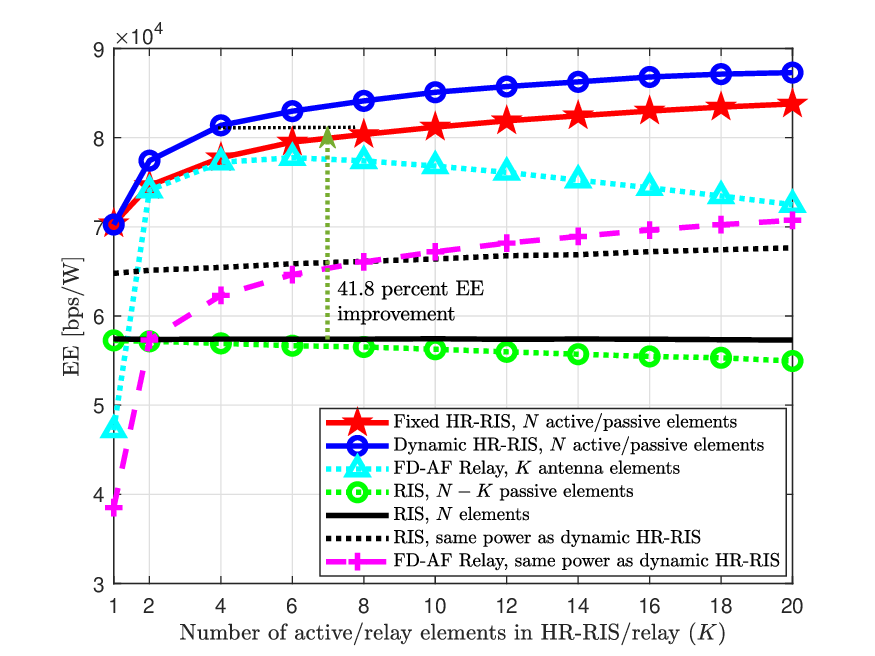}
    }
	\caption{SE and EE of HR-RISs compared to those of conventional RISs and FD-AF relaying.}
	\label{fig_rate}
\end{figure}

To demonstrate the potential of HR-RISs to significantly improve the performance of wireless communications systems, we compare them with conventional RISs and relaying in terms of SE and EE in Fig. \ref{fig_rate}. Specifically, the following schemes are considered:
\begin{itemize}
    \item The proposed fixed and dynamic HR-RISs with $N$ elements, where $K$ elements are active and the remaining $N-K$ elements are passive.
    \item Conventional RISs with $N-K$ or $N$ reflecting elements.
    \item FD-AF relaying with $K$ relay elements employing both analog and digital SI suppression techniques.
\end{itemize}
To maximize the SE, the fully/semi-passive beamforming matrices of the RISs/HR-RISs, respectively, are obtained based on alternating optimization (AO) \cite{nguyen2021hybrid}, while the relay precoder is obtained based on the well-known singular value decomposition.
 
We assume that the BS and MS are equipped with $8$ and $2$ antennas, respectively, and are located at the origin $(0,0)$ meter (m) and $(100,0)$ m in a 2D Cartesian coordinate system, respectively. The HR-RIS, relay, and RIS are assumed to be deployed at the same position, at $(95,1)$ m. We assume that the channel between the BS and HR-RIS is dominated by the line-of-sight link, while that between the HR-RIS and MS follows the Rayleigh fading model \cite{wu2019towards}, with path loss exponents of $2.0$ and $2.8$, respectively \cite{wu2019intelligent}.
The SI channels of the elements of the relay and HR-RIS follow mutually independent and identically distributed complex zero-mean Gaussian distributions with variances of $-95$ dB and $-70$ dB \cite{Bharadia14}, respectively. Furthermore, we ignore the direct BS-MS channel due to the significant propagation loss and blockage between them.

The numerical results are obtained for $N=100$, $K \in [1,20]$ under the assumption of perfect CSI. Two-bit resolution phase shifters are adopted. The transmit power of the BS and the active HR-RIS and relay elements are set to $20$ dBm and $0$ dBm, respectively, while the received noise power is $-170$ dBm/Hz \cite{wu2019towards}. The total power consumption of the system is the sum of the transmit powers of the BS and the HR-RIS/relay, as well as the total circuit power consumption at the BS and the HR-RIS/RIS/relay, see \cite{nguyen2021hybrid} for details.

Fig. \ref{fig_rate}(a) reveals that replacing a few passive elements of the conventional RIS with active elements (making it an HR-RIS), causes only a small loss in passive reflection gain, while the gains provided by the active elements are remarkable. Furthermore, for small numbers of active elements (i.e., small $K$), the HR-RIS achieves significant improvements in both SE and EE. For example, with only four active elements, the dynamic HR-RIS achieves a $42.8$ percent higher SE and a $41.8$ percent higher EE compared with the conventional RIS with the same number of elements (i.e., $100$ elements).

We now compare the performance of the HR-RIS to that of FD-AF relaying. In general, a RIS needs to be very large to achieve a better performance than a relay \cite{bjornson2019intelligent, wu2019intelligent}. Interestingly, the HR-RIS offers better or similar SE and EE performance to the FD relay for small and moderate $K$ (e.g., for $K \leq 10$ in Fig. \ref{fig_rate}(a)). With a large $K$, FD-AF relaying achieves a slightly higher SE than the HR-RIS, but it comes at the cost of very low EE as shown in Fig. \ref{fig_rate}(b), due to the excessively high power consumed by numerous active relay elements.

We further compare the systems aided by the relaying, RISs, and HR-RISs having the same total power consumption. To this end, we increase the transmit power at the BS of the RIS system by the same amount of power as consumed for active processing at the HR-RIS. By contrast, since an FD-AF relay with numerous elements has much higher circuit power consumption than an HR-RIS, we reduce the transmit power at the BS of the former system by the same amount of this circuit power difference. Fig.\ \ref{fig_rate} shows that if the three considered schemes have the same total power consumption, it is difficult for RISs and FD-AF relays to outperform HR-RISs.

\section{Challenges and Future Perspectives}
\subsection{Sustainable HR-RIS Hardware and Implementation}
One challenge in designing HR-RISs is the requirement of low-cost and low-resolution components, including PAs, RF chains, and phase shifters. In general, high-precision components allow for continuous tuning of the phase shifts and amplitudes of the elements to improve the system performance. However, this also leads to a complicated design, high hardware cost, and unaffordable computational complexity, which considerably limits the practical implementation of HR-RISs. Fortunately, the active relaying gain of HR-RISs can to a large extent compensate for the performance degradation due to limited-precision elements, but this has neither been demonstrated with practical prototype, nor thoroughly investigated theoretically. 

To be able to meet the power budget required for energy-sustainable operation and maintenance of HR-RISs, a promising solution is to adopt ultra-low loss switching elements and very low-power digital electronics. Alternatively, green and renewable energy sources can be integrated into HR-RISs given their abundance in most practical deployment scenarios to further improve the energy supply side, alleviating the need for a dedicated power budget \cite{KisseleffOJCOMS20}. 


\subsection{Optimal Configuration and Resource Allocation}
Centralized optimization has been the primary approach for the phase control and resource allocation in RIS-aided communication systems \cite{RenzoJSAC20}. For mobile networks with multiple HR-RISs assisting the communications to a large number of MSs, the computational complexity and the required system resources of centralized solutions might be deemed to be impractical due to the large amounts of control signaling overhead and data to be exchanged between the central processing unit and the BSs/HR-RISs. Hence, distributed solutions performing parallel computations for beamforming, phase control, HR-RIS configuration, and spectrum usage shall be considered.

In addition, joint optimization of beamforming at the BS, MS, and HR-RIS and the scheduling of the MSs is needed to fully exploit the active beamforming gains facilitated by HR-RISs for enhancing the system performance. Such an approach, however, invokes large numbers of continuous and binary optimization variables, intractable design problems, and hybrid analog-digital architectures for the HR-RISs, causing high computational complexity and operational power consumption. Thus, low-complexity design methods have to be developed.

\subsection{Artificial Intelligence-Empowered HR-RISs}
Due to the non-convex nature and the high-dimensional optimization variables of the HR-RIS design problems, the computational complexity needed for solving joint resource allocation problems is exceedingly high. Sophisticated signal processing algorithms are often adopted to obtain effective sub-optimal solutions. Therefore, the application of artificial intelligence (AI)-based designs becomes a necessity to deliver near-optimal solutions with low latency/complexity. Besides, data-driven approaches hold great potential to directly perform channel estimation and localization based on the received data without requiring pilot signals. Nevertheless, existing supervised learning-based approaches are generally not robust to the time-varying and highly dynamic nature of wireless channels in HR-RIS-based systems, which might require frequent retraining of the model, leading to higher capital and operating expenditures. 

\subsection{Emerging Applications}
HR-RISs combining the advantages of both relays and RISs can be beneficial in various wireless architectures and applications. For instance, in mmWave and THz-based systems, the passive reflection and active relaying gains offered by HR-RISs can not only compensate for the considerable path loss and maintain wireless connectivity, but also increase the number of propagation paths, improving the accuracy of sensing and localization. Furthermore, HR-RISs can be applied to cell-free massive MIMO systems, where access points can be replaced with HR-RISs to reduce the cost of network deployment and power consumption. In this regard, the deployment of multiple HR-RISs provides the macro diversity needed to improve coverage and channel capacity enhancement. Of course, the more HR-RISs are incorporated into a network, the more difficult the design of control, signal processing, and synchronization algorithm becomes, and further research is needed to obtain viable solutions.

Besides, the notion of HR-RISs is expected to have a huge impact on the mobile edge computing (MEC), where active elements can be incorporated to either strengthen the reflected channels for offloading computation tasks from the MSs to the MEC server or can be equipped with computing capabilities to aid sharing the workload of the MSs. However, the feasibility of MEC for distant MSs strongly depends on the total energy consumption at the HR-RISs and MSs for data
generation, storage, circuit operation power, transmission, processing, and computation. Another apparent hurdle is keeping the energy consumption and hardware cost of HR-RISs low, which makes edge computing for MSs more difficult. Therefore, a thorough analysis and proper design of the energy-aware MEC for HR-RIS-aided communication systems is indispensable.


\section{Conclusion}
This paper presented the novel concept of HR-RIS-aided wireless communications, a promising unification of conventional RISs and relays, to facilitate the creation of an enhanced intelligent and reconfigurable environment. This is motivated by the major challenges for incorporating RIS in wireless networks, that include the finite-resolution phase shifters and the purely passive reflection. HR-RISs offer a huge performance improvement in terms of SE and EE, providing ubiquitous access and extended coverage. We have highlighted their great potential as a key enabling technology for semi-passive beamforming, channel estimation, and localization in wireless communications systems. We have also pointed out several opportunities and critical challenges that require further investigation to develop this exciting research area and to make the HR-RIS vision practically feasible.


\bibliographystyle{IEEEtran}
\bibliography{IEEEabrv,Ref}

\vspace{0.5cm}
\noindent {\bf Nhan Thanh Nguyen} received the Ph.D. degree from Seoul National University of Science and Technology. He is currently with University of Oulu, Finland.

\vspace{0.5cm}
\noindent {\bf Jiguang He} received the D.Sc. degree from the University of Oulu, Finland, where he currently works as a Postdoctoral Researcher. 

\vspace{0.5cm}
\noindent {\bf Van-Dinh Nguyen}  received Ph.D. degree from Soongsil University. He is currently a Research Associate at University of Luxembourg. 


\vspace{0.5cm}
\noindent {\bf Henk Wymeersch} received the Ph.D. degree from Ghent University, Belgium. He is currently a Professor with the Department of Electrical Engineering, Chalmers University of Technology, Sweden. 


\vspace{0.5cm}
\noindent {\bf Derrick Wing Kwan Ng} [F’21] received his Ph.D. in the ECE department of The University of British Columbia. He is currently a Senior Lecturer and the Scientia Fellow at the University of New South Wales, Australia.

\vspace{0.5cm}
\noindent {\bf Robert Schober} [F’10] received  the Ph.D. degree from Friedrich-Alexander University of Erlangen-Nuremberg, Germany. Since January 2012 he is an Alexander von Humboldt Professor and the Chair for Digital Communication at FAU. 


\vspace{0.5cm}
\noindent {\bf Symeon Chatzinotas}  is Full Professor/Chief Scientist and Co-Head of the SIGCOM Research Group at SnT, University of Luxembourg. 

\vspace{0.5cm}
\noindent {\bf Markku Juntti} [F’20] received his Dr.Sc. degree from University of Oulu, Finland, where he has been a professor since 2000.

\end{document}